\documentclass[aps,prc,twocolumn,showpacs,preprintnumbers,
               nofootinbib,float,superscriptaddress,longbibliography]{revtex4-1}

\usepackage{amsmath,graphicx,float,csquotes,mdframed,appendix,url}
\usepackage[colorlinks=true, pdfstartview=FitV, linkcolor=red, citecolor=blue, urlcolor=blue]{hyperref}
\usepackage{xcolor}
\usepackage{amssymb}

\def\snn{\sqrt{s_\mathrm{NN}}}
\def\Glauber{\textsc{3d-glauber}}
\def\MUSIC{\textsc{music}}
\def\UrQMD{\textsc{urqmd}}
\def\iEBEMUSIC{\textsc{iebe-music}}
\def\GlauberMUSICUrQMD{\textsc{3d-glauber+music+urqmd}}

\begin{document}

\title{The 3D structure of anisotropic flow in small collision systems\\ at the Relativistic Heavy Ion Collider}

\author{Wenbin Zhao}
\affiliation{Department of Physics and Astronomy, Wayne State University, Detroit, Michigan 48201, USA}

\author{Sangwook Ryu}
\affiliation{Department of Physics and Astronomy, Wayne State University, Detroit, Michigan 48201, USA}

\author{Chun Shen}
\affiliation{Department of Physics and Astronomy, Wayne State University, Detroit, Michigan 48201, USA}
\affiliation{RIKEN BNL Research Center, Brookhaven National Laboratory, Upton, NY 11973, USA}

\author{Bj\"orn Schenke}
\affiliation{Physics Department, Brookhaven National Laboratory, Upton, NY 11973, USA}

\begin{abstract}
We present (3+1)D dynamical simulations of asymmetric nuclear collisions at the Relativistic Heavy Ion Collider (RHIC). Employing a dynamical initial state model coupled to (3+1)D viscous relativistic hydrodynamics, we explore the rapidity dependence of anisotropic flow in the RHIC small system scan at 200 GeV center of mass energy. We calibrate parameters to describe central $^3$He+Au collisions and make extrapolations to d+Au and p+Au collisions. Our calculations demonstrate that approximately 50\% of the $v_3(p_T)$ difference between the measurements by the STAR and PHENIX Collaborations can be explained by the use of reference flow vectors from different rapidity regions. This emphasizes the importance of longitudinal flow decorrelation for anisotropic flow measurements in asymmetric nuclear collisions, and the need for (3+1)D simulations.
We also present results for the beam energy dependence of particle spectra and anisotropic flow in d+Au collisions.
\end{abstract}

\maketitle

\section{Introduction}
Ultra-relativistic collisions of heavy ions are expected to create nucleus-sized droplets of  quark-gluon  plasma (QGP), providing a unique opportunity to study the properties of nuclear matter at extreme densities and temperatures.
Precise measurements of various flow observables performed at the Relativistic Heavy Ion Collider (RHIC) and the Large Hadron Collider (LHC) together with the successful descriptions by hydrodynamic calculations have revealed that the created QGP fireball behaves like a nearly perfect fluid with very small specific shear viscosity~\cite{Heinz:2013th, Gale:2013da, Shen:2015msa, Shen:2020mgh}. Over the past ten years, experiments and theorists have employed multiple techniques to assess whether such QGP droplets are also formed in smaller collisions, such as p+A, p+p, and even $\gamma^*$+A collisions~\cite{ATLAS:2021jhn,Zhao:2022ayk,Shen:2022daw} (see~\cite{Dusling:2015gta,Loizides:2016tew,Schlichting:2016sqo,Nagle:2018nvi,Schenke:2019pmk,Schenke:2021mxx} for reviews), and indeed, striking features of collective expansion have been observed in high-multiplicity events of the small collision systems studied at RHIC~\cite{PHENIX:2017xrm,PHENIX:2018lia} and the LHC~\cite{Li:2012hc,Dusling:2015gta,Nagle:2018nvi}.

At RHIC, collisions of proton (p), deuteron (d), and helium-3 ($^3$He) projectiles on gold (Au) targets were proposed to discern whether ``flow-like"  patterns  are  indeed geometry driven and possibly attributable to mini QGP droplet formation.  The PHENIX Collaboration~\cite{PHENIX:2018lia} has reported a large set   of elliptic ($v_2$) and triangular ($v_3$) azimuthal  anisotropy  coefficients for p+Au, d+Au, and $^3$He+Au systems. PHENIX data clearly shows the hierarchy of $v_2$ and $v_3$ expected based on the differences in initial geometry between the three systems.  

These collective features can be quantitatively reproduced by (2+1)D viscous hydrodynamic calculations~\cite{Habich:2014jna,Shen:2016zpp}, which translate the initial spatial anisotropies into final momentum anisotropies of produced hadrons via the collective expansion of the bulk matter. In contrast, calculations  based  solely on initial-state correlations in the color glass condensate framework~\cite{Mace:2018vwq,Mace:2018yvl} are ruled out by the PHENIX data as they predict the opposite hierarchy of $v_2$ between the systems. Recently, the STAR Collaboration has also reported measurements of the differential elliptic flow and triangular flow coefficients in these three small systems~\cite{Lacey:2020ime,STAR:2022pfn}. STAR preliminary results show that the observed $v_3(p_T)$ are system independent, different from the PHENIX published data. 
In \cite{Lacey:2020ime,STAR:2022pfn} sub-nucleon fluctuations are emphasized to play an important role in the initial geometry and modify the expectation for the hierarchy between the flow coefficients in the different systems.

An important difference between the PHENIX and STAR analyses is the use of different pseudo-rapidity ranges in the two-particle correlation measurements that yield the azimuthal anisotropy coefficients. PHENIX measures the two-particle correlations between the backward rapidity ($-3.9<\eta<-3.1$) and mid-rapidity regions ($-0.35<\eta<0.35$) with the event plane method~\cite{PHENIX:2018hho,PHENIX:2018lia}. In contrast, STAR employs only the mid-rapidity ($-0.9<\eta<0.9$) region, using the scalar-product method~\cite{Lacey:2020ime,STAR:2022pfn}. It is already well known that in small asymmetric systems, longitudinal decorrelations are significant \cite{CMS:2015xmx} and boost-invariance is strongly broken~\cite{Bozek:2015swa,Schenke:2016ksl,Shen:2020jwv,Jiang:2021ajc,Wu:2021hkv, Zhao:2022ayk}. More recent calculations also indicate that pre-equilibrium  evolution, either in the weak \cite{Schenke:2019pmk} or strong coupling \cite{Romatschke:2015gxa} limit, has a significant effect on the small and short-lived systems. 

In this work, we include the effects of pre-equilibrium evolution and perform full (3+1)D hydrodynamic simulations, which we argue are essential when seeking meaningful quantitative comparisons to experimental data.

More specifically, we employ a (3+1)D framework combining a dynamic initial state with hydrodynamics and hadronic transport~\cite{Shen:2022oyg} to explore the collective features in small asymmetric systems. The framework has been shown to provide a unified and quantitative description of identified hadron production in $\gamma^*$-nucleus, proton-proton, proton-nucleus, and nucleus-nucleus collisions from center of mass energies of a few GeV to several TeV~\cite{Shen:2022oyg,Zhao:2022ayk}. 
Here we present detailed results on $dN_{ch}/d\eta$, mean transverse momentum ($\left<p_T\right>$), and anisotropic flow in p+Au, d+Au, and $^3$He+Au collisions in this (3+1)D hybrid hydrodynamic framework. 

This paper is organized as follows: Section \ref{sec:model} briefly introduces the \GlauberMUSICUrQMD{} 
model for the description of the p/d/$^3$He+Au systems. Section \ref{sec:results} presents and discusses results from the \GlauberMUSICUrQMD{} model on the $dN_{ch}/d\eta$, $\left<p_T\right>$, and collective flow.  Section \ref{sec:summary} concludes this paper with a summary.

\bigskip
\section{The theoretical framework}\label{sec:model}
In this paper, we employ the \GlauberMUSICUrQMD{} hybrid  model within the open-source \iEBEMUSIC{} framework \cite{iEBEMUSIC} to study paeticle spectra and collective flow observables in p+Au, d+Au, and $^3$He+Au collisions at $\snn=200$\,GeV, and for a range of lower beam energies in d+Au collisions.
The hybrid model uses a 3D Monte-Carlo (MC) Glauber initial condition to dynamically deposit energy, momentum, and net baryon densities into the evolving fluid system as the two colliding nuclei are crossing each other~\cite{Shen:2017bsr, Shen:2022oyg}. The collective expansion of the QGP fireball and the evolution of the conserved net-baryon current are described by a (3+1)D viscous hydrodynamic model \MUSIC~\cite{Schenke:2010rr, Schenke:2010nt, Paquet:2015lta, Denicol:2018wdp, MUSIC}.  As the QGP expands and becomes more dilute in the hadronic phase, the fluid dynamic description is switched to a microscopic hadron cascade model, \UrQMD~\cite{Bass:1998ca, Bleicher:1999xi, UrQMD}, to simulate the subsequent evolution and dynamic decoupling of the hadronic matter.

\begin{figure*}[t]
  \includegraphics[scale=0.9]{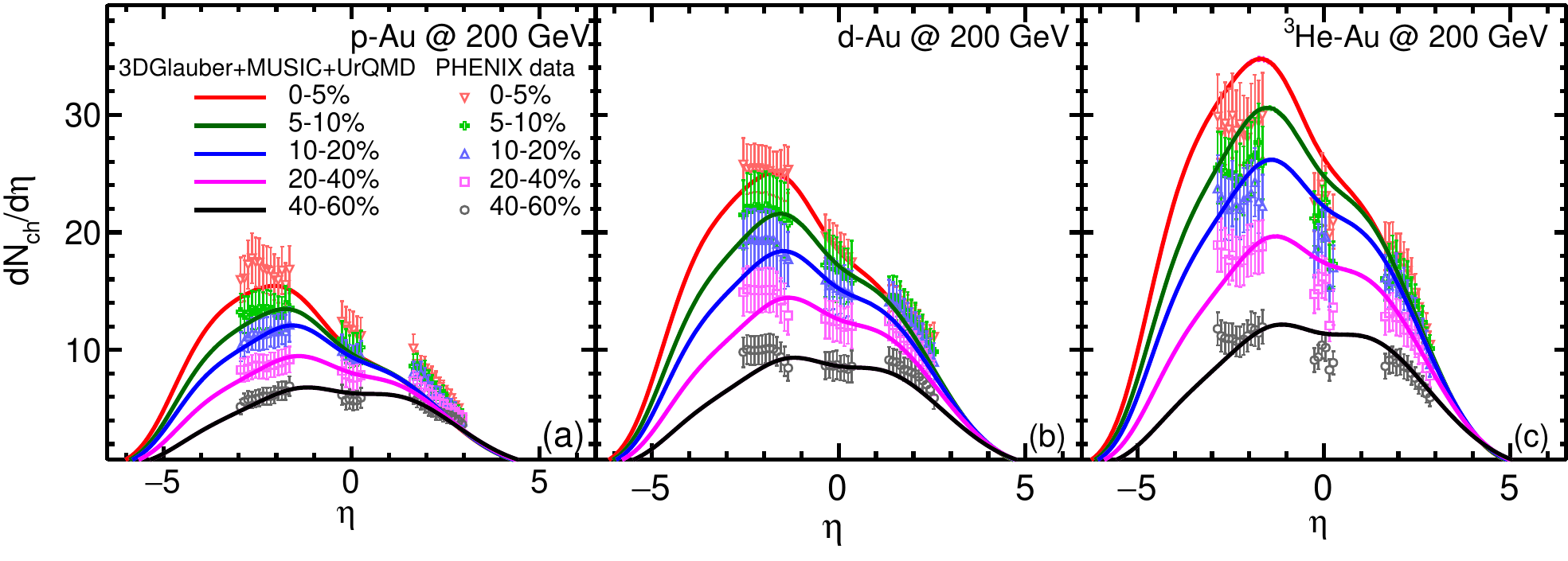}
  \caption{(Color online) The charged hadron pseudo-rapidity distributions $dN_{\rm ch}/d\eta$ 
in various multiplicity classes of p+Au, d+Au and
$^3$He+Au collisions at $\sqrt{s_{N}}$ = 200 GeV
 from the \GlauberMUSICUrQMD{} simulations, compared to experimental data from the PHENIX Collaboration~\cite{PHENIX:2018hho}.}
  \label{fig:dnchdeta}
\end{figure*}
\begin{figure}[t]
  \includegraphics[scale=0.4]{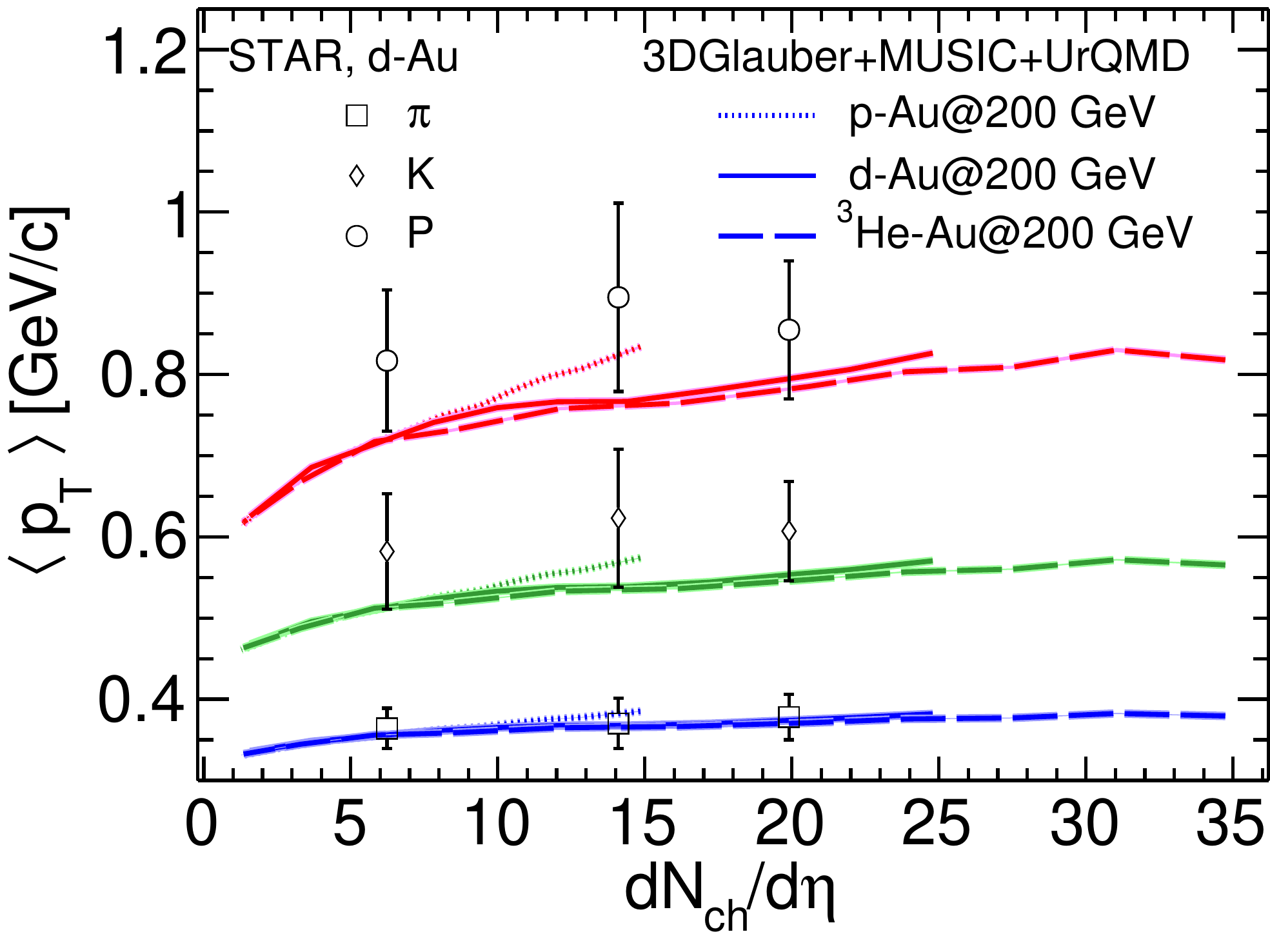}
  \caption{(Color online) Identified  particle  mean  transverse momenta $\left< p_T\right>$ as functions of charged hadron multiplicity in p+Au (solid lines), d+Au (dotted lines) and $^3$He+Au (dashed lines) collisions from the \GlauberMUSICUrQMD{} framework. The d+Au results are compared to experimental data from the STAR Collaboration \cite{STAR:2008med}.
  }
  \label{fig:meanpt}
\end{figure}
More specifically, the dynamical initial condition is simulated by the 3D Monte-Carlo Glauber model on an event-by-event basis~\cite{Shen:2017bsr,Shen:2022oyg}, and the space-time and momentum distributions of the initial energy-momentum tensor and net baryon  charge current are provided by the classical string deceleration model~\cite{Bialas:2016epd,Shen:2017bsr}. The transverse positions of valence quarks and soft partonic cloud in the \Glauber{} model are sampled from a 2D Gaussian, $\exp\left[ - \frac{x^2 + y^2}{2}/B^2_{G} \right]$, with a  transverse nucleon width parameter $B_{G}=5$ GeV$^{-2}$, corresponding to a nucleon width of $w = 0.44$\,fm.
After the collision, the deposited energy density distribution has a Gaussian profile in the transverse plane with a width of $w_q = 0.2$\,fm. We parameterize the average rapidity loss function of the valence quarks and the soft partonic cloud with their incoming rapidity $y_\mathrm{init}$ in the collision pair rest frame as~\cite{Shen:2018pty,Shen:2022oyg},
\begin{equation}
    \langle y_\mathrm{loss} \rangle (y_\mathrm{init}) = A y_\mathrm{init}^{\alpha_2} [\tanh(y_\mathrm{init})]^{\alpha_1 - \alpha_2}.
    \label{eq:ylossMean}
\end{equation}
As shown in Fig.~\ref{fig:dnchdeta}, when choosing the three parameters values $A=1.32, \alpha_1 = 1.8$, and $\alpha_2=0.34$, we can reproduce the measured pseudo-rapidity distributions of charged hadrons for p+Au, d+Au, and $^3$He+Au collisions at $\snn=200$\,GeV. The spatial topography of  the  deuteron’s  two-nucleon  system  is  obtained  from sampling the Hulthen wave function~\cite{PHENIX:2013jxf}, and fluctuating $^3$He configurations come from results of Green's function Monte Carlo calculations using the AV18+UIX model interaction~\cite{Carlson:1997qn}.  The detailed implementation of this initial condition model is discussed in Ref.~\cite{Shen:2022oyg}.

To incorporate the effects of pre-hydrodynamic flow, we include a finite transverse initial velocity motivated by the blast-wave model~\cite{Schnedermann:1993ws, Teaney:2000cw, Heinz:2004qz}. We assume the initial flow is radial and its strength is characterized by the transverse flow rapidity,
\begin{equation}
    \eta_\perp = \alpha^{\rm pre-flow} r,
    \label{eq:etaPerp}
\end{equation}
where $r$ is the radial distance from the string center. The $\alpha^{\rm pre-flow}$ parameter controls the strength of the pre-hydrodynamic flow. Here we set $\alpha^{\rm pre-flow} = 0.15$ to fit the $v_n(p_T)$ in $^3$He+Au at 200 GeV. We will explore the effects of pre-hydrodynamic flow on hadronic observables in p+Au collisions in Appendix~\ref{sec:Appendix-B}.

The produced strings from individual nucleon-nucleon collisions are treated as dynamical source terms for the hydrodynamic evolution~\cite{Okai:2017ofp, Shen:2017ruz,Shen:2017bsr,Du:2018mpf,Shen:2022oyg},
\begin{eqnarray}
    \partial_\mu T^{\mu\nu} &=& J^\nu \\
    \partial_\mu J_B^\mu &=& \rho_B.
\end{eqnarray}
Here we use a crossover equation of state ({\tt NEOS-BQS}) for the QCD matter at finite chemical potentials, that is constructed using recent lattice data~\cite{Borsanyi:2011sw, Borsanyi:2013bia, Ding:2015fca, Bazavov:2017dus, Monnai:2019hkn}. We employ the strangeness neutrality condition, $n_s=0$, and for simplicity set the net electric charge-to-baryon density ratio to $n_Q/n_B=0.4$~\cite{Monnai:2019hkn}, a value most appropriate for large nuclei.
The specific shear and bulk viscosities are parametrized as
\begin{equation}
    \frac{\eta T}{e+P}=(\eta/s)_{\rm norm}\left[1+(\eta/s)_{\rm slope}\left(\frac{\mu_B}{\mu_{B, \rm scale}}\right)^\alpha\right]
    \label{eq:shear}
\end{equation}
\begin{equation}
    \frac{\zeta T}{e+P}=(\zeta/s)_{\rm norm}{\rm exp}\left[-\left(\frac{T-T_{\rm peak}}{T_{{\rm width},\lessgtr}}\right)^2\right],
    \label{eq:bulk}
\end{equation}
where $T_{\rm width,<} = 0.015$\,GeV for $T<T_{\rm peak}$, and $T_{\rm width, >} = 0.1$\,GeV for $T> T_{\rm peak}$ with $T_{\rm peak}=0.17$ GeV and $(\zeta/s)_{\rm norm}=0.08$. For the specific shear viscosity, we set $(\eta/s)_{\rm norm}=0.13$, $(\eta/s)_{\rm slope}=1.0$, $\mu_{B, \rm scale}=0.6$ GeV, and $\alpha =1.5$.
The shear stress tensor and bulk viscous pressure are evolved with the Denicol-Niemi-Moln\'{a}r-Rischke (DNMR) theory with spatial gradients up to the second order~\cite{Denicol:2012cn, Denicol:2018wdp}. For simplicity, the effect of charge diffusion is neglected in this work. These transport coefficients are chosen to reproduce the mean transverse momentum and the differential flow in central $^3$He+Au collisions at 200 GeV. In addition, the $\mu_B$-dependence of $\eta/s$ is informed by calibrating flow observables in Au+Au collisions from 7.7 GeV to 200 GeV \cite{Sangwook:bulk}.
Compared to Ref.~\cite{Shen:2022oyg}, we use smaller sizes for nucleons and hotspots, which are consistent with the sub-nucleonic structure constrained by a recent Bayesian analysis \cite{Mantysaari:2022ffw}. The small hotspot size ($w_q = 0.2$\,fm) and inclusion of pre-hydrodynamic flow require a non-zero QGP bulk viscosity in the hydrodynamic evolution, which was neglected in \cite{Shen:2022oyg}.

In the \iEBEMUSIC{} framework, the  Cooper-Frye  particlization of the fluid cells is performed  on a hyper-surface with a constant energy density of $e_{\rm sw}=$0.50 GeV/fm$^3$ using the Cornelius algorithm \cite{Huovinen:2012is} and the open-source code package \textsc{iSS} \cite{Shen:2014vra, iSS_code}. The non-equilibrium shear and bulk viscosity corrections to the local equilibrium distribution function are taken from Grad's moments method with multiple conserved charges~\cite{Sangwook:bulk}.
The produced hadrons are then fed into the hadron cascade model, \UrQMD, for further scatterings and decays until kinetic freeze-out is achieved dynamically. 

\begin{figure}[t]
  \includegraphics[scale=0.450]{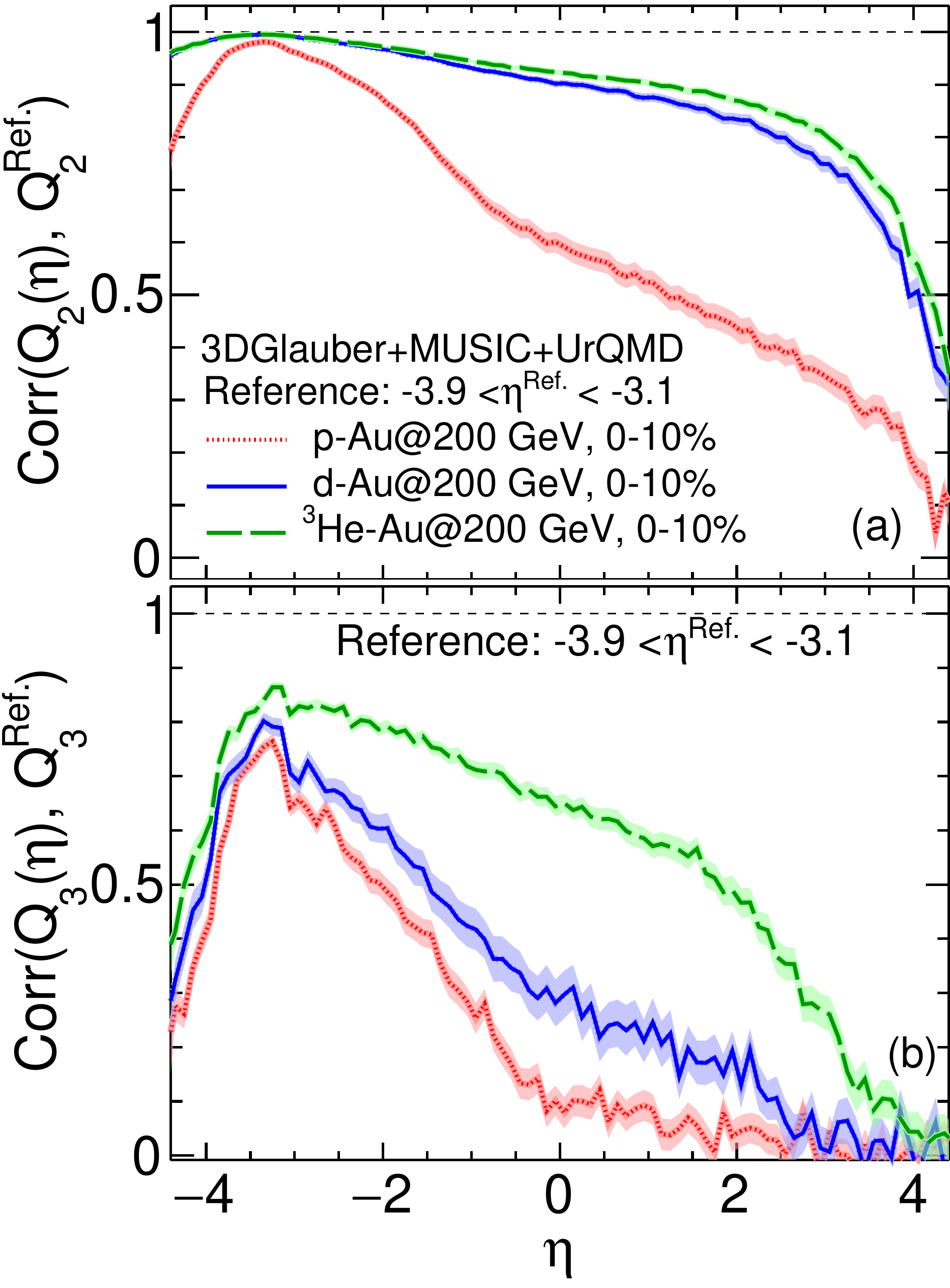}
  \caption{(Color online) The correlation between the flow vectors as a function of pseudo-rapidity $\eta$ with the reference vector calculated within $-3.9<\eta^{\rm Ref.}<-3.1 $ in p+Au, d+Au and $^3$He+Au systems from the \GlauberMUSICUrQMD{} simulations.}
  \label{fig:correlation}
\end{figure}
\begin{figure*}[t]
  \includegraphics[scale=0.9]{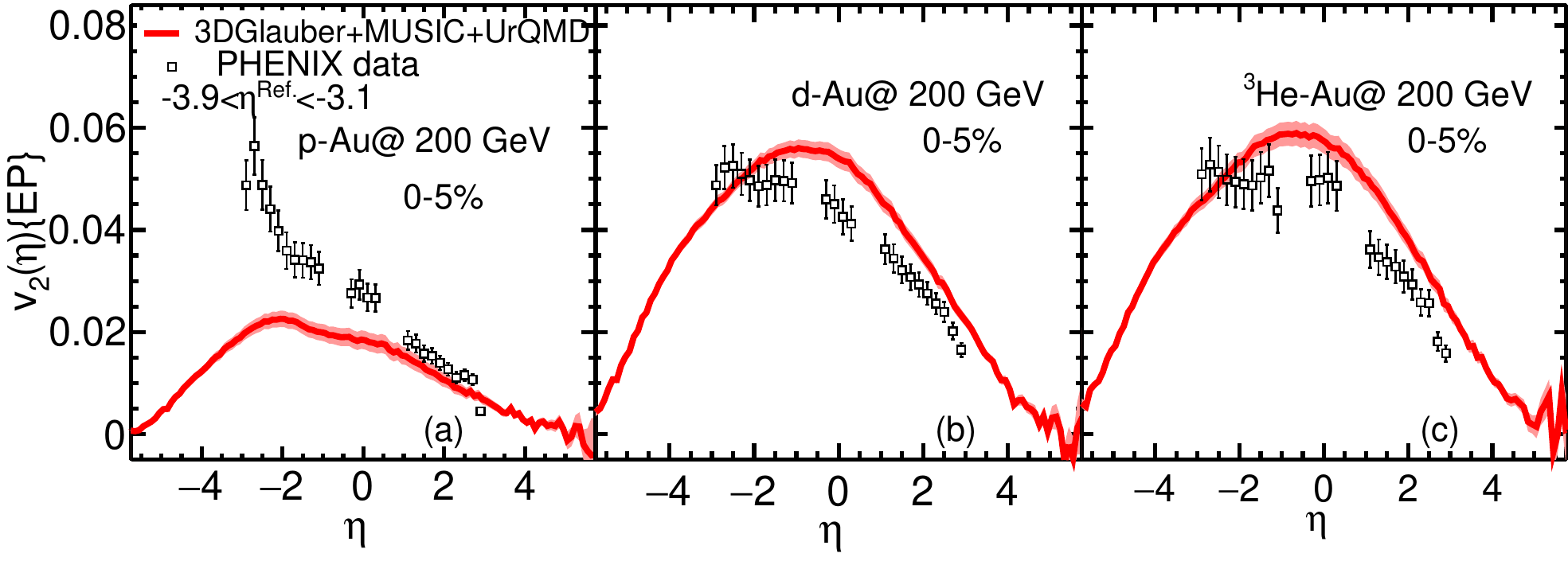}
  \caption{(Color online)  Elliptic flow $v_2$ as a function of pseudo-rapidity ($\eta$) in high-multiplicity 0-5\% central p+Au, d+Au and $^3$He+Au collisions at $\snn = 200$ GeV from the \GlauberMUSICUrQMD{} framework.  The results are compared to the experimental data from the PHENIX Collaboration~\cite{PHENIX:2018hho}.}
  \label{fig:v2eta}
\end{figure*}
\begin{figure*}[t]
  \includegraphics[scale=0.9]{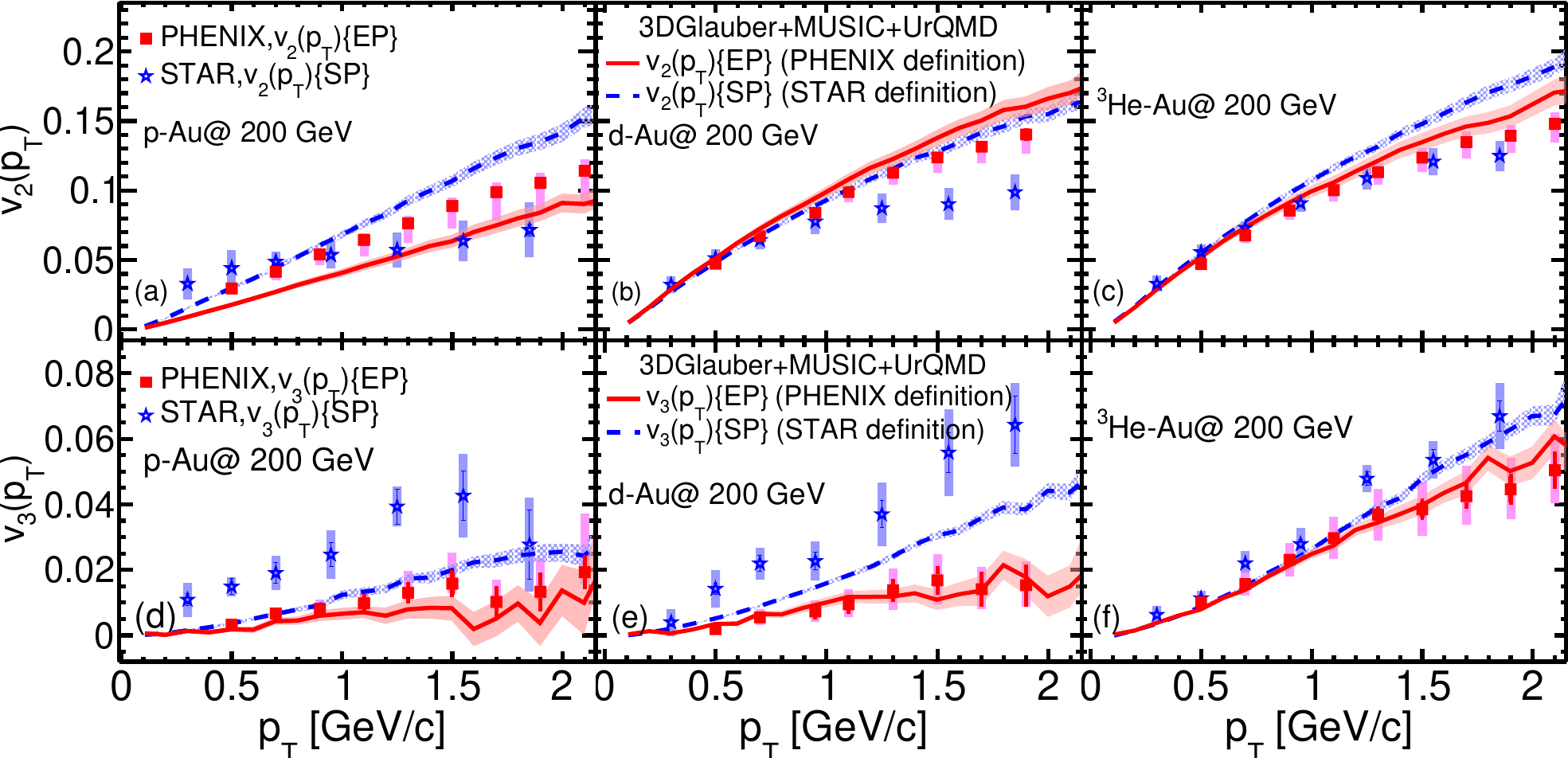}
  \caption{(Color online)  The anisotropic flow $v_n(p_T)$ as a function of $p_T$ in central p+Au, d+Au and $^3$He+Au collisions computed from the \GlauberMUSICUrQMD{} framework. The results are compared to the experimental data from the PHENIX and STAR Collaborations~\cite{PHENIX:2018lia,Lacey:2020ime}. The PHENIX data are all 0-5\% and STAR data are 0-2\%, 0-10\%, and 0-10\% for p+Au, d+Au, and $^3$He+Au collisions, respectively.}
  \label{fig:v2pt2pc}
\end{figure*}

\bigskip
\section{RESULTS}\label{sec:results}
In this section, we report results for observables, including charged hadron pseudo-rapidity distributions, mean transverse momentum, and differential flow harmonics for the p+Au, d+Au, and $^3$He+Au collision systems at RHIC energies.  

\subsection{Charged Hadron Multiplicity Rapidity Distribution and Average Transverse Momentum}
Figure~\ref{fig:dnchdeta} shows the charged hadron pseudo-rapidity distributions in different centrality bins for p+Au, d+Au, and $^3$He+Au collisions at $\snn=200$\,GeV. We note that all parameters were tuned to reproduce observables in $^3$He+Au collisions.
Keeping all the model parameters fixed, our model's extrapolations provide a reasonable description of the experimental data from the PHENIX Collaboration \cite{PHENIX:2018hho} for the p+Au and d+Au systems.
In particular, the increase of the asymmetry around midrapidity from peripheral to central collisions is well captured as a consequence of the increasing number of participant nucleons from the Au target towards central collisions.
Here we have used the charged hadron multiplicity in the pseudo-rapidity range $-3.9 < \eta < -3.1$ (in the Au-going direction) to determine the collision centrality bins, which is consistent with the method used by PHENIX \cite{PHENIX:2018hho}.
As pointed out in Ref.~\cite{Shen:2022oyg}, the correlation of particle production between forward and mid-rapidity is crucial to reproduce the centrality dependence of charged particles in these asymmetric collision systems.
This is highlighted in Fig.~\ref{fig:dNchdetacentrality} in Appendix~\ref{sec:Appendix-A}, where we show that using different centrality definitions generates different shapes for $dN_{ch}/d\eta$.

Within the hydrodynamic paradigm, the mean transverse momentum of identified hadrons strongly constrains the radial expansion of the fireball. This radial expansion is very sensitive to the bulk viscosity, initial hotspot size, and the strength of pre-hydrodynamic flow. Consequently, the mean transverse momentum provides a strong constraint on $(\zeta/s)(T)$~\cite{Ryu:2015vwa,Ryu:2017qzn}.
In Fig.~\ref{fig:meanpt} we show the identified particle mean transverse momentum as a function of charged hadron multiplicity in p+Au, d+Au, and $^3$He+Au collisions at 200 GeV and compare to the STAR measurements for d+Au collisions~\cite{STAR:2008med}. Using the temperature-dependent bulk viscosity $(\zeta/s)(T)$ in Eq.\,\eqref{eq:bulk} with $(\zeta/s)_{\rm norm}$ = 0.08, the calculation shows good agreement with the STAR data in d+Au collisions \cite{STAR:2008med}. This indicates that our hybrid model simulations generate proper amounts of radial flow and the transverse expansion has been well constrained.

\begin{figure*}[t]
  \includegraphics[scale=0.9]{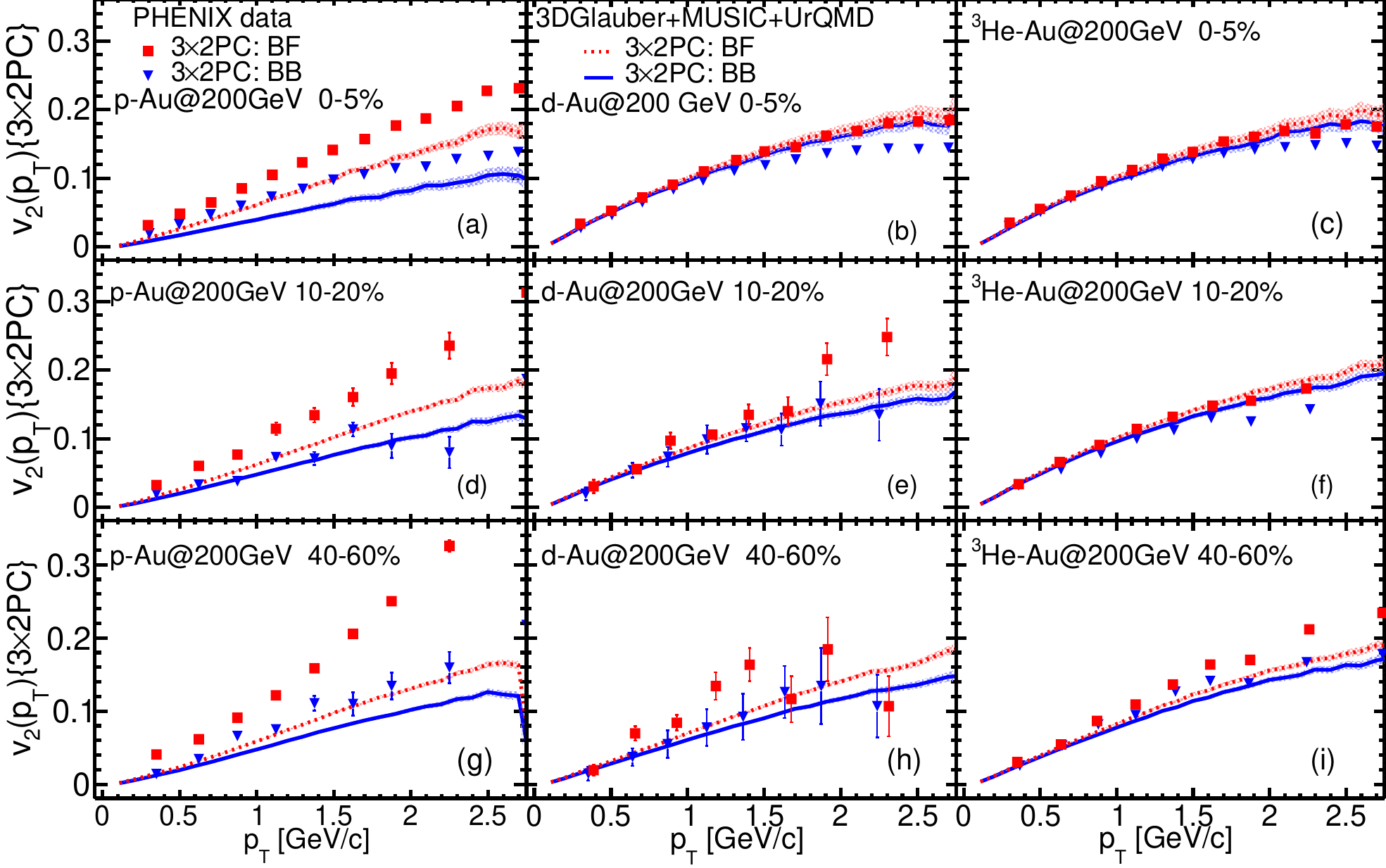}
  \caption{(Color online) The differential elliptic flow $v_2(p_T)$ in $^3$He/d/p+Au collisions at $\snn=200$ GeV determined by the 3$\times$2PC method using pseudo-rapidity ranges corresponding to the FVTXS-CNT-FVTXN (BF) and BBCS-FVTXS-CNT (BB) detector combinations. The model results are compared to experimental data from the PHENIX  Collaboration~\cite{PHENIX:2021ubk,PHENIX:2022nht}}
  \label{fig:v2pt3pcthree}
\end{figure*}

The flow harmonics describe the momentum anisotropy of the final produced hadrons. They are commonly measured using the two-particle correlation method with a large pseudo-rapidity gap between the particles to suppress non-flow correlations~\cite{Bilandzic:2010jr,Bilandzic:2013kga}. 
The presence and size of the gap will affect the measured value of $v_n\{2\}$ if the correlation between the flow vectors depends on the pseudo-rapidity separation of two particles.
Therefore, we calculate the $\eta$-dependent correlations between the flow vectors $Q_n$ defined by 
\begin{equation}\label{eq:correlation}
{\rm Corr}\left(Q_n(\eta),~Q^{\rm Ref.}_n\right)=\frac{\left\langle Q^*_n(\eta)Q_n^{\rm Ref.}\right\rangle}{\sqrt{\left\langle \vert Q_n(\eta)\vert^2\right\rangle}\sqrt{\left\langle \vert Q_n^{\rm Ref.} \vert^2\right\rangle}},
\end{equation}
to characterize the longitudinal decorrelation of flow vectors. Here, $Q_n(\eta)$ is the $n$-th order flow vector defined in the $\eta$ bin of interest, and $Q_n^{\rm Ref.}$ is the $n$-th order reference flow vector within the assigned reference $\eta$ range.
Figure~\ref{fig:correlation} shows the correlation ${\rm Corr}\left(Q_n(\eta),~Q^{\rm Ref.}_n\right)$ of the second and third order flow vectors with their reference flow vectors calculated within $-3.9<\eta<-3.1$ in central 0-10\% p+Au, d+Au, and $^3$He+Au systems, following PHENIX's detector coverage~\cite{PHENIX:2018lia}.

The values of ${\rm Corr}\left(Q_n(\eta),~Q^{\rm Ref.}_n\right)$ for $|\eta| < 0.35$ show the strength of flow vector correlations relevant for the PHENIX measurement of $v_n\{2\}$ \cite{PHENIX:2018lia}. For elliptic flow, we find that the correlation drops slowly with pseudo-rapidity and is approximately 0.9 for $|\eta| < 0.35$ in d+Au and $^3$He+Au collisions. However, the flow vector correlation drops quite rapidly in p+Au collisions, reaching $\sim0.6$ in the $|\eta| < 0.35$ bin, which indicates that the $v_2\{2\}$ measurements could depend strongly on the rapidity regions used in the analysis. For the triangular flow coefficients, the flow correlations are significantly below unity for all three systems and drop more rapidly with $\eta$ compared to the case of $n=2$. This result demonstrates that it is non-trivial to extract the fireball's initial triangularity hierarchy from $v_3\{2\}$ measurements. Because the boost-invariant (2+1)D simulations can not capture these longitudinal flow correlations, their results for triangular flow in (p,d,$^3$He)+Au collisions are highly idealized and contain substantial theoretical uncertainties.

The three panels of Figure~\ref{fig:v2eta} show the elliptic flow $v_2(\eta)$ as a function of pseudo-rapidity in the 0–5\% centrality bin of $\snn=200$\,GeV p+Au, d+Au, and $^3$He+Au collisions, respectively. Following the same method as used by the PHENIX Collaboration \cite{PHENIX:2018hho}, we calculate the second flow harmonic, $v_2(\eta)$, using the event plane (EP) method with the event plane determined in the range $-3.9<\eta<-3.1$, corresponding to using the Beam Beam Counter (BBC) in the Au-going direction, and with centrality defined by the charged hadron multiplicity measured in the same range. The flow harmonics for the event plane method are calculated using 
\begin{equation}\label{eq:vnep}
v_n(\eta, p_T)\{\mathrm{EP}\} = \frac{\langle \cos(n(\phi(\eta, p_T)-\Psi^\mathrm{Ref.}_n) \rangle}{R(\Psi^\mathrm{Ref.}_n)},
\end{equation}
where $\phi(\eta,p_T)$ is the azimuthal angle of  particles of interest for specific $\eta$ and/or $p_T$ bins, $\Psi^{\rm Ref.}_n$ is the $n$-th order azimuthal
reference event plane, and ${\rm R(\Psi^{Ref.}_n)}$ is the resolution of $\Psi^{\rm Ref.}_n$. Following the PHENIX measurements~\cite{PHENIX:2017xrm,PHENIX:2017nae,PHENIX:2018hho}, the event plane resolution is estimated by:
\begin{equation}\label{eq:resolution}
{\rm R(\Psi^{Ref.}_n)}=\sqrt{\frac{\langle\frac{Q_{nA}}{\vert Q_{nA}\vert}\frac{Q^{*}_{nB}}{\vert Q_{nB}\vert}\rangle\langle\frac{Q_{nA}}{\vert Q_{nA}\vert}\frac{Q^{*}_{nC}}{\vert Q_{nC}\vert}\rangle}{\langle\frac{Q_{nB}}{\vert Q_{nB}\vert}\frac{Q^{*}_{nC}}{\vert Q_{nC}\vert}\rangle}}
\end{equation}
with the $n$th-order flow vector,
\begin{equation}
    Q_{n} = \sum_{k} e^{i n \phi_k},
\end{equation}
with the $\sum$ summing over particles in one event.
The $Q_{nA,B,C}$ vectors are defined in the $-3.9<\eta<-3.1$ (south  BBC, Au-going side),  $-3.0<\eta< -1.0$ (south Forward Silicon Vertex (FVTX)) and $\vert\eta\vert<0.35$ (central - CNT), respectively.
The \GlauberMUSICUrQMD\ model gives a reasonable description of the $\eta$-dependent elliptic flow in d+Au and $^3$He+Au systems. Both the experimental data and our model simulations have an increasing flow coefficient $v_2$ towards backward rapidity up to $\eta = -3.0$ in the Au-going direction. At $\eta<-3.0$, our model predicts a decrease of $v_2(\eta)$. The $v_2(\eta)$ of $^3$He+Au is slightly larger than that of d+Au collisions because of the longer fireball lifetime in the hydrodynamic phase in $^3$He+Au collisions.
For the p+Au system, the PHENIX data shows a pronounced increase of the $v_2(\eta)$ towards negative rapidities, especially for $\eta<-2.0$. Our model fails to describe the $v_2(\eta)$ for $\eta<$ 1.0 in p+Au collisions. We expect $v_2(\eta)$ in p+Au collisions to be smaller than in the other systems because of the stronger decorrelation and in p+Au collisions, compared to d+Au and $^3$He+Au collisions, as shown in Fig.~\ref{fig:correlation}. Also the shorter lifetime in p+Au collisions can lead to smaller $v_2(\eta)$.

Figure~\ref{fig:v2pt2pc} shows the computed $p_T$-differential flow $v_n(p_T)$ $(n=2,3)$ for charged hadrons compared to experimental data from the PHENIX and STAR Collaborations. Here, we adopt the same methods and kinematic cuts as employed in the PHENIX or STAR analyses, respectively. Specifically, when comparing to the PHENIX data~\cite{PHENIX:2018lia}, we select collision events for 0-5\% centrality using the number of charged hadrons in $\eta \in [-3.9, -3.1]$.
The $v_n(p_T)\{\rm EP\}$ (PHENIX) for charged hadrons in the midrapidity region covering $\vert\eta\vert<0.35$ (central CNT) are calculated with the event plane method by Eq.\,(\ref{eq:vnep}), where the second-order event plane is determined in the Au-going direction, in the pseudo-rapidity range $-3.0<\eta< -1.0$ (south FVTX) in p/d+Au and $-3.9<\eta<-3.1$ (south  BBC) in $^3$He+Au collisions, and the third order event planes are determined in the range of the south  BBC  for all systems \cite{PHENIX:2018lia}. The event plane resolutions are calculated with the three-subevent method, which correlates measurements in the south BBC, south FVTX, and central CNT arms~\cite{PHENIX:2018lia,PHENIX:2018hho} as given explicitly in Eq.\,(\ref{eq:resolution}).
The centrality classes are defined using the multiplicity of charged hadrons in the pseudo-rapidity range $-3.9<\eta<-3.1$. 

To compare with the experimental data from the STAR Collaboration \cite{Lacey:2020ime,STAR:2022pfn}, we select the 0-2\% most central p+Au collisions and 0-10\% most central d+Au and $^3$He+Au collisions, based on the charged hadron multiplicity in $\eta \in [-0.9, 0.9]$.
The STAR flow measurements imposed a rapidity gap $|\Delta \eta| > 1$ for charged particles in $\eta \in [-0.9, 0.9]$. For convenience we compute the two-particle flow coefficients using the scalar-product (SP) method with particles of interest (POIs) and reference particles (RPs) selected from two sub-events within $-0.9<\eta<-0.5$ and $0.5<\eta<0.9$, respectively. We confirmed that this approximates STAR's method very well.

As mentioned above, we tune parameters to fit observables, including $v_2(p_T)$ and $v_3(p_T)$, in $^3$He+Au collisions, and then predict observables in p+Au and d+Au collisions.  Fig.\,\ref{fig:v2pt2pc} shows that the \GlauberMUSICUrQMD{} model gives an overall good description of the PHENIX $v_n(p_T)$ data for d+Au and $^3$He+Au collisions. We underestimate the $v_2(p_T)$ in p+Au collisions by about 20-30\%, possibly because of a too large longitudinal flow decorrelation in the model.
Compared to the STAR elliptic flow, our model provides a reasonable description for $p_T < 1$ GeV for (d, $^3$He)+Au collisions, but underestimates the data for p+Au collisions. Our calculation overestimates the STAR $v_2(p_T)$ at higher $p_T$. 
The STAR $v_3(p_T)$ measurements show a much weaker system size dependence than the model, which gives a good description of $v_3(p_T)$ in $^3$He+Au but significantly underestimates $v_3(p_T)$ in (p, d)+Au collisions.

Both in the model calculations and experimental data, the $v_3(p_T)$ with the STAR definition (stars) are systemically larger than those determined using the PHENIX definition (squares). In our calculations, this difference is mainly caused by the different magnitudes of the longitudinal decorrelation of flow vectors of $v_3$ between the different pseudo-rapidity bins. That is, the decorrelation between the flow vectors at $-3.9<\eta<-3.1$ and $-0.35<\eta<0.35$ is greater than that between $-0.9<\eta<-0.5$ and $0.5<\eta<0.9$.

However, our model still underestimates the STAR data of $v_3(p_T)$ in p+Au and d+Au collisions, indicating that the decorrelations present in the model do not fully account for the differences between the STAR and PHENIX results.
We leave a more systematic calibration of fitting both STAR and PHENIX data for future studies using a Bayesian framework. This study should shed more light on the role of any residual non-flow in the experimental data.

Recently, the PHENIX Collaboration measured anisotropic flow coefficients using the $3\times$2PC method, whose two-particle azimuthal correlations are constructed with three different sets of pairs~\cite{PHENIX:2021ubk,PHENIX:2022nht}. Taking the $p_T$-integrated flow vector $Q_{nA,B,C}$, and the $p_T$-differential flow vector $q_{nA,B,C}(p_T)$ of particles of interest (using finite $p_T$ bins) from different sub-events $A$, $B$, and $C$ that indicate different pseudo-rapidity bins, the differential $v_n(p_T)$ of the particles of interest in the single sub-event $C$ can be determined as
\begin{equation}
v_n^C(p_T) = \sqrt{ \frac{c_n^{AC}(p_T)c_n^{BC}(p_T)}{C_n^{AB}} }.
\end{equation}
where the  $p_T$-differential $c_n(p_T)$ and $p_T$-integrated $C_n$ coefficients are defined as 
\begin{align}
C_n^{AB} &= \left\langle Q_{nA} Q^{*}_{nB} \right\rangle, \\
c_n^{AC}(p_T) &=  \left\langle Q_{nA} q^{*}_{nC}(p_T)\right\rangle, \\
c_n^{BC}(p_T) &=  \left\langle Q_{nB} q^{*}_{nC}(p_T)\right\rangle.
\end{align}
Here $\left\langle...\right\rangle$ represents event averaging and taking the real part of the correlator.

Figure \ref{fig:v2pt3pcthree} shows the differential elliptic flow $v_2(p_T)\{3\times2{\rm PC}\}$  at mid-rapidity $\vert\eta\vert<0.35$ in $^3$He+Au, d+Au, and p+Au systems, respectively. Following the PHENIX measurements~\cite{PHENIX:2021ubk,PHENIX:2022nht}, we calculate the $v_2(p_T)$ from the 3×2PC method with two combinations, ``$-3.0<\eta<-1.0,~-0.35<\eta<0.35~{\rm and}~1.0<\eta<3.0$", FVTXS-CNT-FVTXN, detector combination (BF) and ``$-3.9<\eta<-3.1,~-3.0<\eta<-1.0~{\rm and}~-0.35<\eta<0.35$", BBCS-FVTXS-CNT, detector combination (BB).

The \GlauberMUSICUrQMD{} model reproduces the $v_2(p_T)\{3\times2{\rm PC}\}$ in d+Au and $^3$He+Au collisions quite well up to 60\% in centrality. The model calculations produce a slowly decreasing trend in $v_2(p_T)$'s magnitude from central to peripheral collisions for both the BF and BB combinations. This trend is caused by the shorter hydrodynamic lifetimes in more peripheral collisions. More importantly, our model quantitatively reproduces the difference between $v_2(p_T)({\rm BF})$ and $v_2(p_T)({\rm BB})$, which suggests that our model predicts the right amount of longitudinal flow decorrelation in d+Au and $^3$He+Au collisions from most central to middle central collisions. Our model results underestimate the measurements in the peripheral collisions, for example in 40-60\% bin, where we expect sizable non-flow correlations in the measurements.

Similar to the previous $v_n\{2\}$ comparison, our calculations underestimate the p+Au $v_2(p_T)$ by 40\% across all centrality bins. Interestingly, our model can reasonably reproduce the difference between $v_2(p_T)({\rm BF})$ and $v_2(p_T)({\rm BB})$ in 0-5\% and 5-10\% central p+Au collisions. Furthermore, this $v_2(p_T)$ difference between BF and BB is larger in p+Au collisions than in (d, $^3$He)+Au collisions, suggesting a large longitudinal flow decorrelation in p+Au collisions, as present in our model.
The PHENIX $v_2(p_T)({\rm BF})$ for 20\% central or more peripheral p+Au collisions show rapid increases at high $p_T$, suggesting sizable residual non-flow correlations.

\begin{figure}[t]
  \includegraphics[scale=0.4]{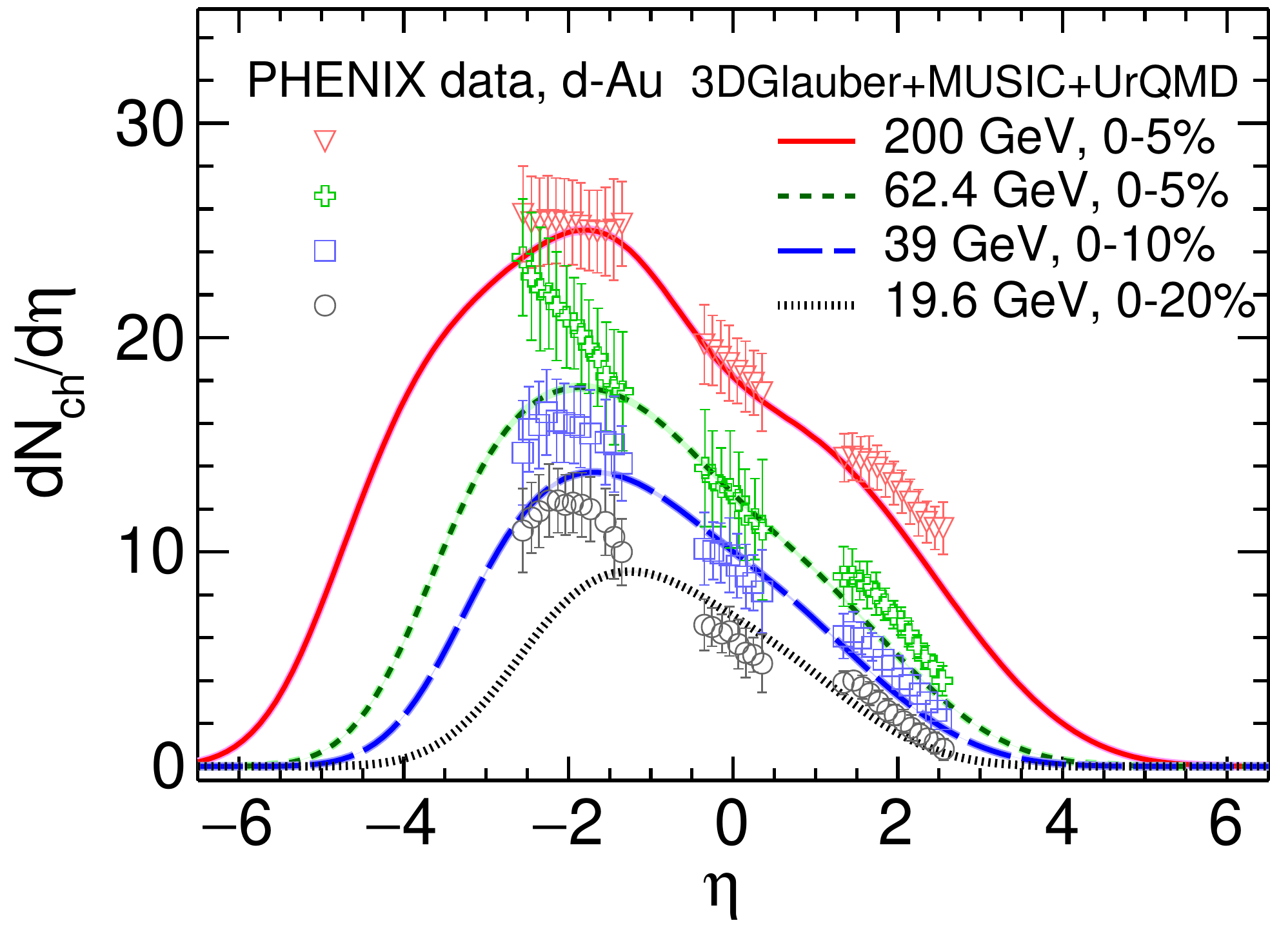}
  \caption{(Color online) The charged hadron pseudo-rapidity distributions $dN_{\rm ch}/d\eta$ of central d+Au collisions at $\snn$ = 200, 62.4, 39 and 19.6 GeV  from the \GlauberMUSICUrQMD{} simulations, compared to experimental data from the PHENIX Collaboration~\cite{PHENIX:2017nae}.}
  \label{fig:dNchdetaBES}
\end{figure}
\begin{figure*}[t]
  \includegraphics[scale=0.9]{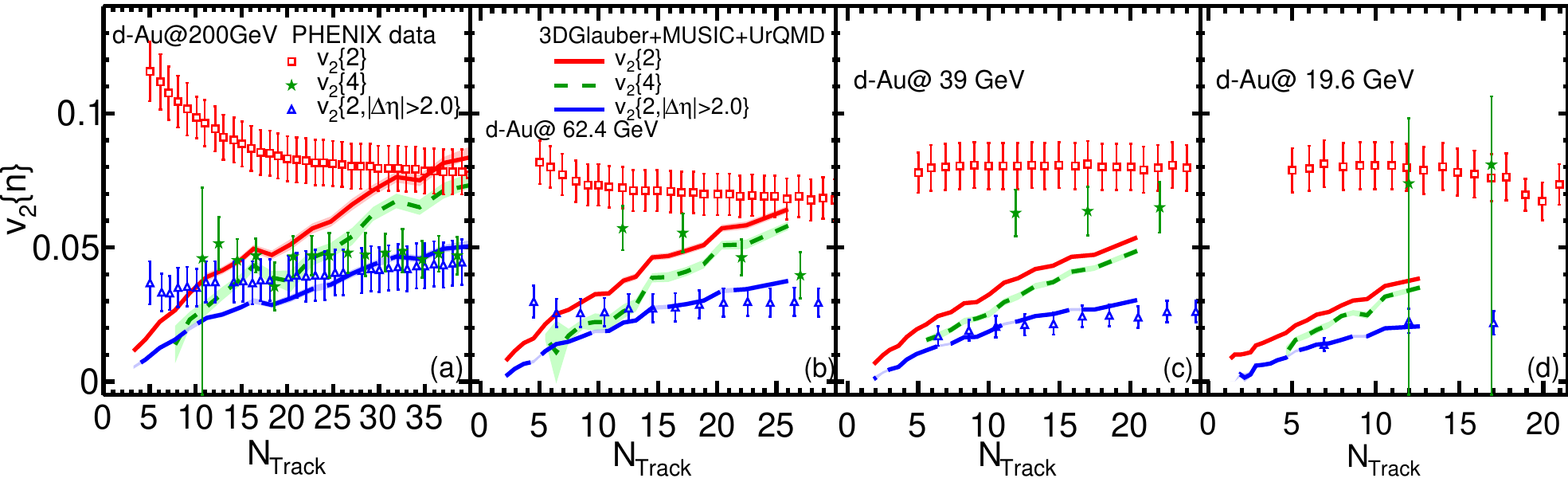}
  \caption{(Color online)  Charged hadron elliptic flow $v_2\{2\}$, $v_2\{4\}$ and $v_2\{2,\vert\Delta\eta\vert>2.0\}$  as a function of charged hadron multiplicity $N_{\rm Track}$ of  d+Au collisions at $\snn = 200$, 62.4, 39 and 19.6 GeV from the \GlauberMUSICUrQMD{} framework.  The results are compared to experimental data from the PHENIX Collaboration~\cite{PHENIX:2017xrm}.}
  \label{fig:v2BES}
\end{figure*}
\begin{figure*}[t]
  \includegraphics[scale=0.9]{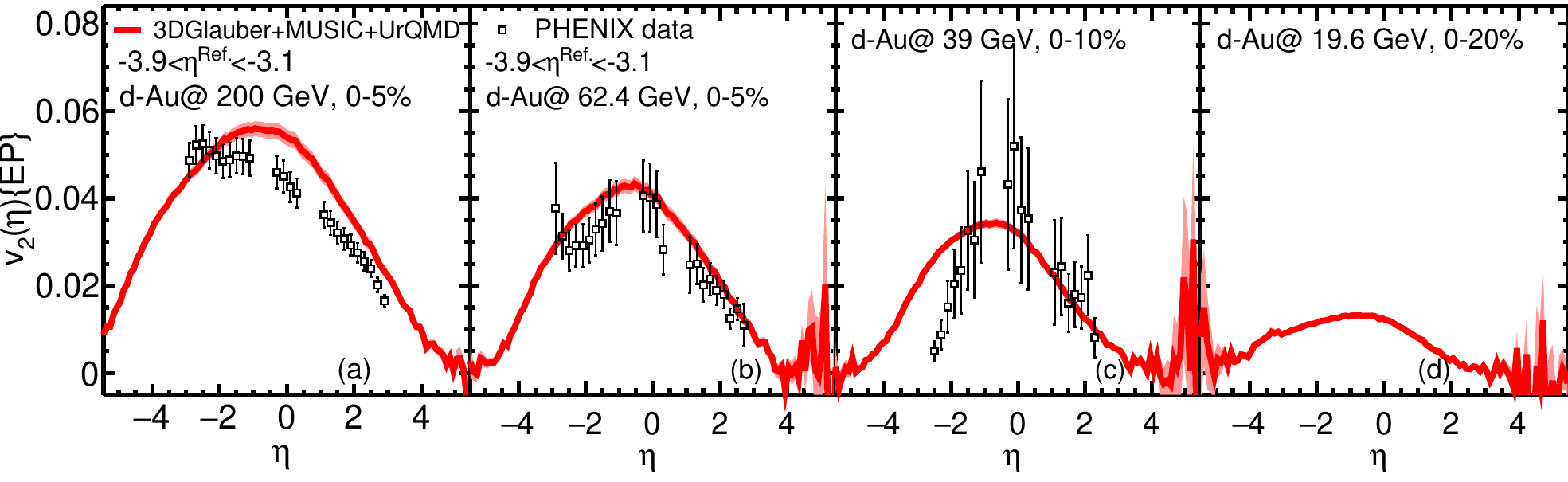}
  \caption{(Color online)  Elliptic flow $v_2$ as a function of pseudo-rapidity in high-multiplicity central  d+Au   collisions at $\snn = 200$, 62.4, 39 and 19.6 GeV from the \GlauberMUSICUrQMD{} framework.  The results are compared to the experimental data from the PHENIX Collaboration~\cite{PHENIX:2017nae}.}
  \label{fig:v2etaBES}
\end{figure*}
\subsection{Results for the d+Au Beam Energy Scan}
In this sub-section, we extrapolate our model to d+Au collisions at lower collision energies. To do so we only reduce the center-of-mass collision energy and keep all model parameters fixed.

Figure \ref{fig:dNchdetaBES} shows our model predictions for the collision energy dependence of the charged hadron pseudorapidity distribution in central d+Au collisions.  Our model gives an overall good description of the asymmetric charged hadron pseudorapidity distributions compared with the PHENIX data from $\snn = $ 200 GeV to 19.6 GeV. The peaks of $dN_{ch}/d\eta$ in the theoretical calculations are shifted more towards midrapidity compared with the PHENIX measurements at low collision energies, suggesting a small overestimation of the rapidity loss in the initial-state model for the lower collision energies.
Compared to the results shown in Ref.~\cite{Shen:2022oyg}, the current $dN_{ch}/d\eta$ at 19.6 GeV is slightly smaller because a higher switching energy density $e_{\rm sw} = 0.5$\,GeV/fm$^3$ is used here for all collision energies. 

Figure \ref{fig:v2BES} shows the model-to-data comparisons for $v_2\{2\}$, $v_2\{4\}$ and $v_2\{2,\vert\Delta\eta\vert>2.0\}$ in d+Au collisions at 200, 62.4, 39, and 19.6 GeV, using the PHENIX FVTX ($1.0<\vert\eta\vert<3.0$) acceptance~\cite{PHENIX:2017xrm}. Here, following the PHENIX measurements~\cite{PHENIX:2017xrm,PHENIX:2018lfu}, the $N_{\rm Track}$ is estimated by $N_{\rm Track} = p_0 \times N_{ch}$, with $p_0=0.67$ being the tracking efficiency and acceptance in the FVTX detector. The $N_{ch}$ is the charged hadron multiplicity within the FVTX acceptance $1<\vert\eta\vert<3$ computed in our model.
With a rapidity gap of $\vert\Delta\eta\vert>2.0$, non-flow correlations can be effectively suppressed. 

We find our calculations agree nicely with the PHENIX $v_2\{2,\vert\Delta\eta\vert>2.0\}$ data for all four collision energies. 
The difference between $v_2\{2\}$ and $v_2\{2,\vert\Delta\eta\vert>2.0\}$ data indicate the contributions from event plane decorrelation and non-flow correlations. In our calculations, the difference between these two correlation observables only comes from the event plane decorrelations. Therefore, the difference between the model and PHENIX $v_2\{2\}$ can serve as an estimation of non-flow correlations in the measurements. 

The elliptic flow measured using the four-particle cumulant method, $v_2\{4\}$, in general, contains fewer non-flow correlations.
Our model results overestimate $v_2\{4\}$ for central d+Au collisions at 200 GeV. For the lower collision energies, our $v_2\{4\}$ underestimates the PHENIX data. At the current stage, it is not clear whether the discrepancy originates from possible residual non-flow correlation in the measurements or different variances in the elliptic flow distributions.

Finally, Figure~\ref{fig:v2etaBES} shows the rapidity-dependent $v_2(\eta)\{\rm EP\}$ in central d+Au collisions at $\snn$ = 200, 62.4, 39 GeV, and 19.6 GeV, calculated using the same event plane method as that in Fig.~\ref{fig:v2eta}. Our model produces results close to the PHENIX measurements at the top three collision energies. Our $v_2(\eta)\{\rm EP\}$ at 19.6 GeV serves as a model prediction.
The $v_2(\eta)\{\rm EP\}$ at all four energies have similar shapes with respect to $dN_{\rm ch}/d\eta$ that decreases with increasing $\eta$ between $0 < \eta < 3$ in the proton-going side, and peaks at the Au-going side. This correlation between $v_2(\eta)$ and $dN_{\rm ch}/d\eta$ reflects that the size of the elliptic flow is proportional to the fireball lifetime in the hydrodynamic phase~\cite{PHENIX:2017nae}.

\bigskip
\section{Summary }\label{sec:summary}
We explored the collectivity in asymmetric (p, d, $^3$He)+Au collisions at 200 GeV and d+Au collisions at 19.6 - 200 GeV within a (3+1)D hybrid theoretical framework. We find that the elliptic flow vector correlations in the PHENIX kinematic regions remain $\sim 0.9$ for central (d, $^3$He)+Au collisions at 200 GeV, ensuring a good signal of the geometric response from initial 2D ellipticity to final-state charged hadron elliptic flow. However, the elliptic flow correlation drops to 0.6 in central p+Au collisions. For triangular flow, the flow correlations are significantly below unity for all three collision systems. The presence of substantial flow decorrelations leads to the conclusion that full (3+1)D simulations are required for any quantitative comparisons to experimental data from these asymmetric collision systems.

With parameters calibrated using the central $^3$He+Au measurements, the \GlauberMUSICUrQMD{} model reproduces the PHENIX two-particle $v_n(p_T)$ in central d+Au collisions, but underestimates the p+Au $v_n(p_T)$ by 30\%. By computing $v_n(p_T)$ using the STAR definitions, we find a 30\% larger $v_3(p_T)$ with the STAR definition compared to the triangular flow from the PHENIX definition. Compared to the STAR measurements, our calculations suggest that approximately 50\% of the difference between the PHENIX and STAR $v_3(p_T)$ originates from the different flow correlations between different rapidity regions.
The remaining discrepancy between our calculations and the experimental data could come from different origins, including pre-hydrodynamic flow and non-flow in the experimental data that is not present in the model. 

We further obtain a reasonable agreement with PHENIX measurements of differential elliptic flow using the $3\times$2PC method in d+Au and $^3$He+Au collisions up to 60\% in centrality. We underestimate the $v_2(p_T)$ in p+Au by approximately 40\%. However, our model quantitatively reproduces the difference between the $v_2(p_T)$ measured using the BF and BB combinations of detectors for all three systems. The larger observed difference between $v_2(p_T)({\rm BF})$ and $v_2(p_T)({\rm BB})$ in p+Au collisions compared to (d, $^3$He)+Au collisions suggests a larger longitudinal flow decorrelation in p+Au collisions, as present in our model.

Lastly, we extrapolate our model calculations to d+Au collisions at 19.6, 39, and 62.4 GeV. With all the model parameters fixed by the $^3$He+Au collisions at 200 GeV, our model predictions give good descriptions of PHENIX measurements of $v_2\{2\}$ with a rapidity gap and the rapidity dependent elliptic flow in d+Au at the lower collision energies.

In summary, our study explicitly demonstrates that longitudinal flow decorrelations play a central role in anisotropic flow measurements in asymmetric nuclear collisions. We conclude that full (3+1)D hybrid simulations are essential for making quantitative comparisons with observables from small system collisions at RHIC.

\bigskip
\section*{Acknowledgements}
We thank Ron Belmont for providing the PHENIX data and Mashhood Munir for estimates of the pre-hydrodynamic flow. We thank Shengli Huang, Ulrich W. Heinz, Jiangyong Jia and Zhengbu Xu for helpful discussions and suggestions.  W.B.Z. is supported by the National Science Foundation (NSF) under grant numbers ACI-2004571 within the framework of the XSCAPE project of the JETSCAPE collaboration. S.R. is supported by the U.S. Department of Energy(DOE) under grant number DE-SC0013460. B.P.S. and C.S. are supported by the U.S. Department of Energy, Office of Science, Office of Nuclear Physics, under DOE Contract No.\,DE-SC0012704 and Award No. DE-SC0021969, respectively. C.S. acknowledges support from a DOE Office of Science Early Career Award. 
This work is in part supported within the framework of the Beam Energy Scan Theory (BEST) Topical Collaboration and under contract number DE-SC0013460.
This research was done using resources provided by the Open Science Grid (OSG) \cite{Pordes:2007zzb, Sfiligoi:2009cct}, which is supported by the National Science Foundation award \#2030508.

\appendix
\section{Centrality Selection Dependence of $dN_{ch}/d\eta$ in Asymmetric Collisions}
\label{sec:Appendix-A}

\begin{figure}[t]
  \includegraphics[scale=0.4]{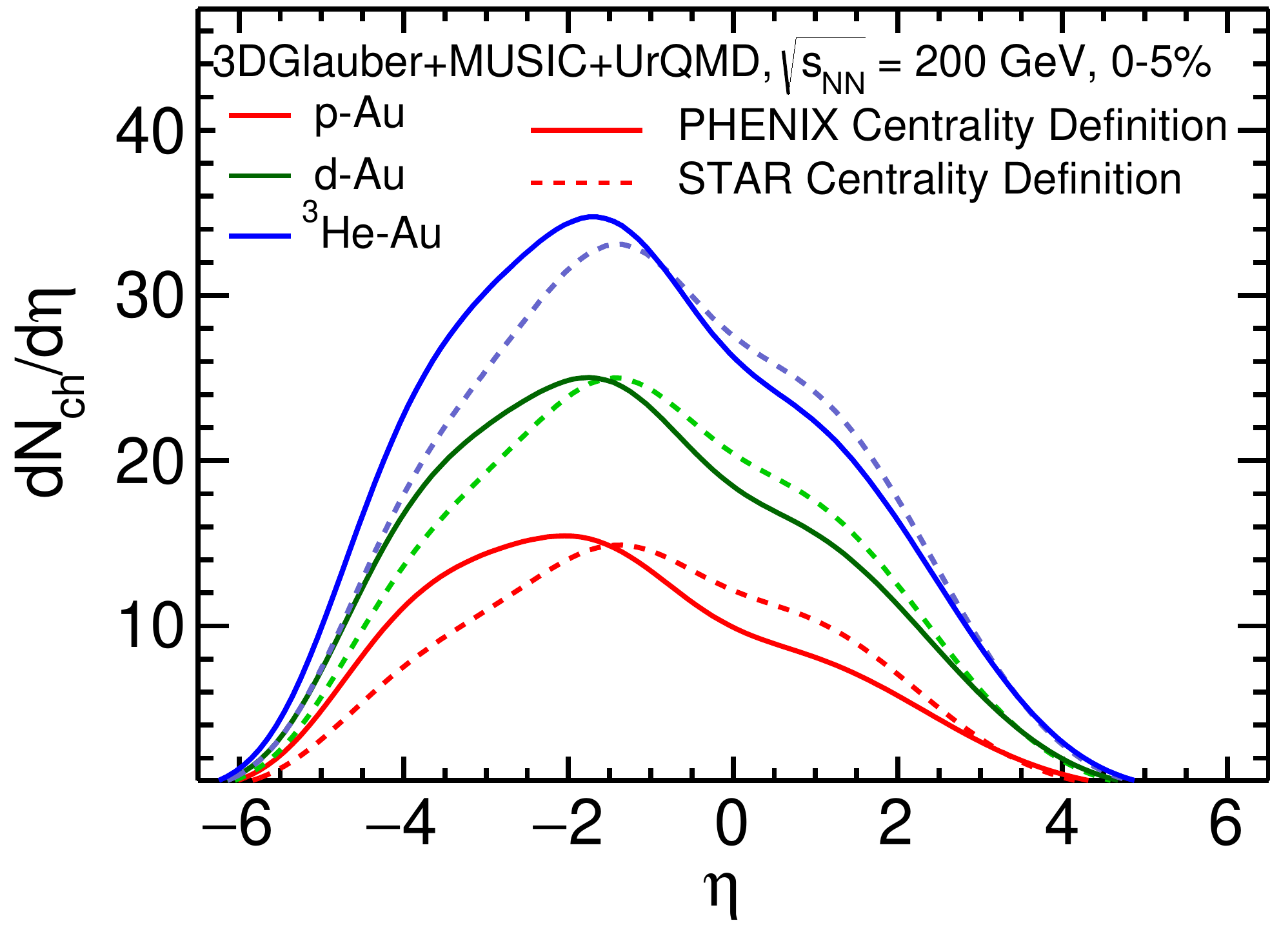}
  \caption{(Color online) The charged hadron pseudo-rapidity distributions $dN_{\rm ch}/d\eta$ with different centrality selections of central p+Au, d+Au and $^3$He+Au collisions at $\snn$ = 200 GeV from the \GlauberMUSICUrQMD{} simulations.}
  \label{fig:dNchdetacentrality}
\end{figure}

We study the effects of different centrality definitions on the charged hadron pseudo-rapidity distributions in 0-5\% central (p, d, $^3$He)+Au collisions at 200 GeV. Figure~\ref{fig:dNchdetacentrality} shows the comparison of the $dN_{ch}/d\eta$ between the PHENIX centrality definition (determined by charged hadron multiplicity within $-3.9<\eta<-3.1$)  and the STAR centrality definition (determined by charged hadron multiplicity within $-0.9<\eta<0.9$). The STAR centrality definition gives a larger yield at mid-rapidity than that using PHENIX's definition, while the central events according to the PHENIX definition have larger yields in the backward (negative) rapidity region. Figure~\ref{fig:dNchdetacentrality} illustrates that it is crucial to adopt the same centrality definition as the experiment for model-to-data comparisons in these asymmetric collisions, emphasizing again the need for full (3+1)d hybrid simulations.

\section{Importance of Pre-Hydrodynamic Evolution for Anisotropic Flow in p+Au Collisions}
\label{sec:Appendix-B}
\begin{figure}[t]
  \includegraphics[scale=0.4]{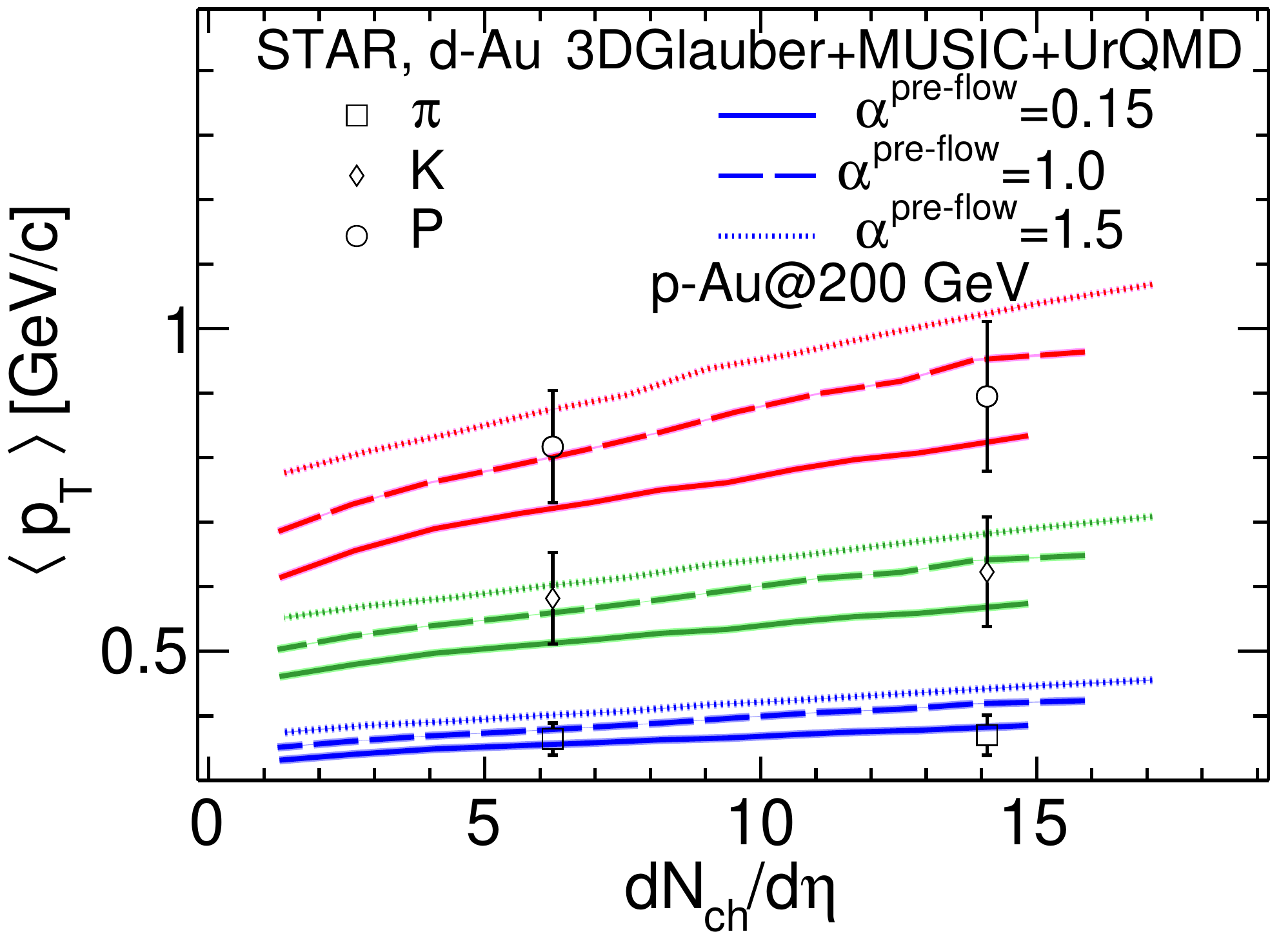}
  \caption{(Color online)  Identified  particle  mean  transverse momenta $\left< p_T\right>$ as functions of charged hadron multiplicity in p+Au  collisions with different values of $\alpha^{\rm pre-flow}$ from the \GlauberMUSICUrQMD{} framework. The results are compared to experimental d+Au data from the  STAR Collaboration \cite{STAR:2008med}.}
  \label{fig:meanptpreflow}
\end{figure}

\begin{figure}[t]
  \includegraphics[scale=0.4]{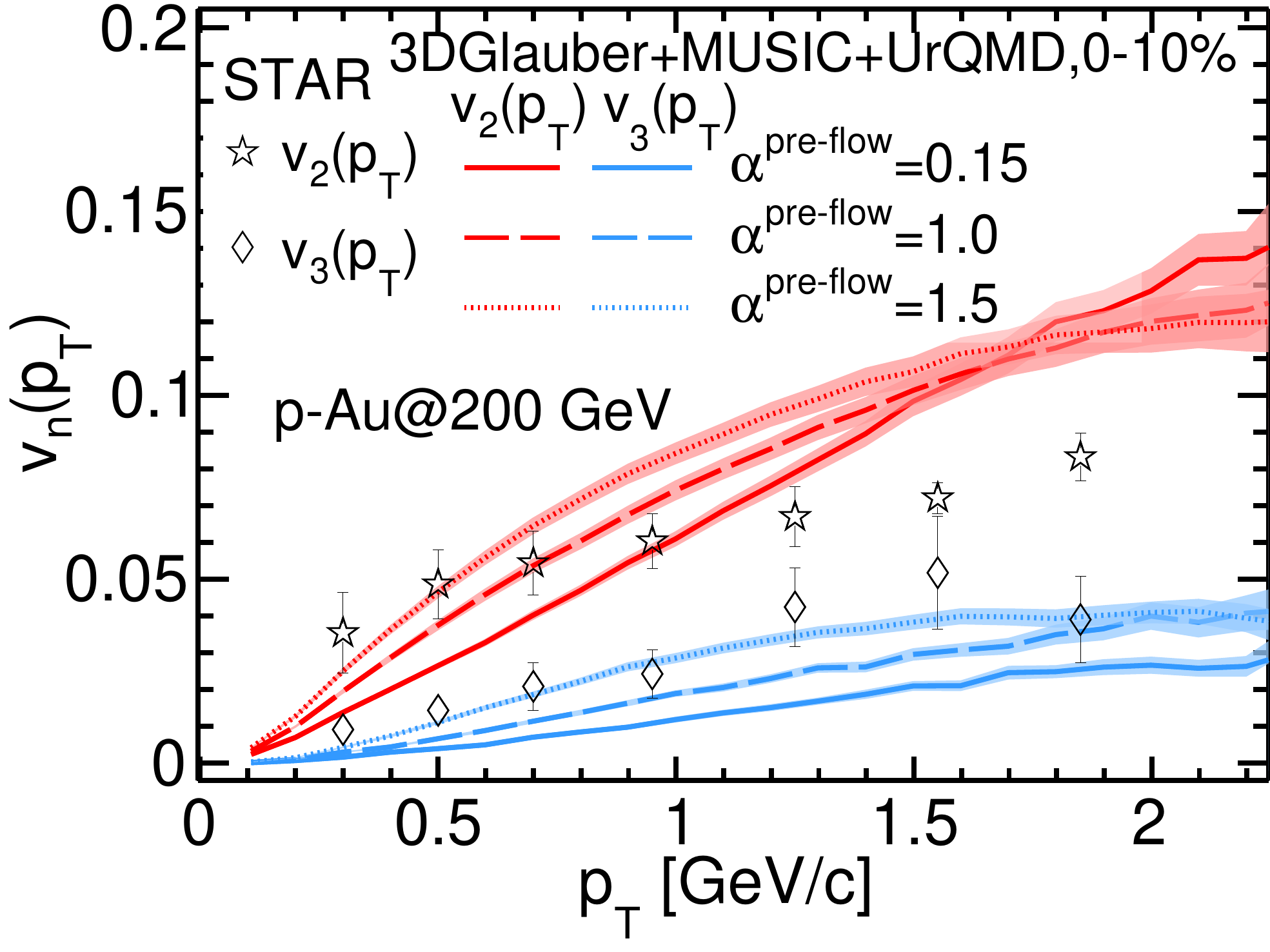}
  \caption{(Color online)  The charged hadron anisotropic flow $v_n(p_T)$ as a function of $p_T$ in central p+Au collisions with different values of $\alpha^{\rm pre-flow}$ computed from the \GlauberMUSICUrQMD{} framework. The results are compared to experimental data from the STAR Collaboration~\cite{Lacey:2020ime}.}
  \label{fig:vnpreflow}
\end{figure}

We investigate the importance of the pre-hydrodynamic flow for hadronic observables in p+Au collisions. Figs.~\ref{fig:meanptpreflow} and~\ref{fig:vnpreflow} show the identified particle mean transverse momenta and the $v_n(p_T)$ ($n=2, 3$) with different values of the $\alpha^{\rm pre-flow}$ defined in Eq.~\eqref{eq:etaPerp} when simulating central p+Au collisions at 200 GeV. Here the $\alpha^{\rm pre-flow}$ parameter controls the strength of the pre-hydrodynamic flow. Figure~\ref{fig:meanptpreflow} shows that a larger pre-hydrodynamic flow leads to a faster transverse expansion, which results in larger $\langle p_T\rangle$ for identified hadrons. The proton's mean $p_T$ shows a stronger sensitivity to $\alpha^{\rm pre-flow}$ than that of pions and kaons.
It's also noteworthy that the faster fireball expansion with a larger $\alpha^{\rm pre-flow}$ shortens the overall fireball lifetime and its longitudinal expansion. Therefore, there is slightly more energy left at mid-rapidity to produce more charged hadrons.

Figure~\ref{fig:vnpreflow} shows that the pre-hydrodynamic flow has significant effects on the charged hadron anisotropic flow in p+Au collisions.
At $p_T \sim 1$\,GeV, the elliptic flow increases by 50\% and triangular flow increases by 300\% when we change from $\alpha^{\rm pre-flow}=0.15$ to $\alpha^{\rm pre-flow}=1.5$. The stronger sensitivity of higher order anisotropic flow to pre-hydrodynamic flow is in qualitative agreement with early results from boost-invariant simulations~\cite{Romatschke:2015gxa}.
With $\alpha^{\rm pre-flow}=1.5$, our simulations can reproduce the STAR $v_3(p_T)$ data but they overestimate the $v_2(p_T)$ for $p_T > 0.6$\,GeV at the same time. This result is in qualitative agreement with results shown in Ref.~\cite{Schenke:2020mbo}. 


\bibliography{references,noInspires}

\end{document}